\documentclass[amssymb,amsmath,twocolumn,superscriptaddress]{aastex63}

\usepackage[caption=false]{subfig}
\usepackage{graphicx}
\usepackage{bm}
\usepackage{amssymb,amsmath}
\usepackage{mathrsfs}
\usepackage{latexsym}
\usepackage{color}
\usepackage[normalem]{ulem}
\usepackage{dcolumn}

\allowdisplaybreaks

\newcommand{\CCA}{\affiliation{Center for Computational Astrophysics, Flatiron Institute, 162 5th Ave, New York, NY 10010, USA}}
\newcommand{\STBR}{\affiliation{Department of Physics and Astronomy, Stony Brook University, Stony Brook NY 11794, USA}}

\definecolor{kcmagenta}{rgb}{0.54, 0.17, 0.88}
\definecolor{darkturquoise}{rgb}{0.0, 0.81, 0.82}

\newcommand{\cn}[1]{{{\textcolor{blue}{\bf [cn]}}}}

\newcommand{\p}{\hat{p}}

\newcommand\DirectRateMed{30.1}
\newcommand\DirectRateHigh{88.9}
\newcommand\DirectRateLow{24.9}
\newcommand\DirectGammaMed{1.1}
\newcommand\DirectGammaHigh{1.9}
\newcommand\DirectGammaLow{2.3}
\newcommand\DirectMmaxMed{42.7}
\newcommand\DirectMmaxHigh{18.6}
\newcommand\DirectMmaxLow{6.4}
\newcommand\DirectAlphaMed{2.3}
\newcommand\DirectAlphaHigh{13.5}
\newcommand\DirectAlphaLow{24.7}
\newcommand\DirectAlphaUpperLim{13.7}
\newcommand\DirectStochRateMed{33.6}
\newcommand\DirectStochRateHigh{90.1}
\newcommand\DirectStochRateLow{27.5}
\newcommand\DirectStochGammaMed{1.0}
\newcommand\DirectStochGammaHigh{1.7}
\newcommand\DirectStochGammaLow{2.5}
\newcommand\DirectStochMmaxMed{42.8}
\newcommand\DirectStochMmaxHigh{20.1}
\newcommand\DirectStochMmaxLow{6.3}

\newcommand\DirectStochAlphaUpperLim{10.1}

\newcommand\DesignDirectRateMed{29.2}
\newcommand\DesignDirectRateHigh{7.2}
\newcommand\DesignDirectRateLow{6.4}
\newcommand\DesignDirectGammaMed{1.3}
\newcommand\DesignDirectGammaHigh{0.3}
\newcommand\DesignDirectGammaLow{0.3}
\newcommand\DesignDirectMmaxMed{44.9}
\newcommand\DesignDirectMmaxHigh{1.2}
\newcommand\DesignDirectMmaxLow{1.2}
\newcommand\DesignDirectAlphaMed{3.2}
\newcommand\DesignDirectAlphaHigh{0.8}
\newcommand\DesignDirectAlphaLow{0.6}
\newcommand\DesignDirectZpeakLowerLim{1.7}
\newcommand\DesignDirectStochRateMed{28.9}
\newcommand\DesignDirectStochRateHigh{8.5}
\newcommand\DesignDirectStochRateLow{7.5}
\newcommand\DesignDirectStochGammaMed{1.2}
\newcommand\DesignDirectStochGammaHigh{0.3}
\newcommand\DesignDirectStochGammaLow{0.3}
\newcommand\DesignDirectStochMmaxMed{45.1}
\newcommand\DesignDirectStochMmaxHigh{1.7}
\newcommand\DesignDirectStochMmaxLow{1.0}
\newcommand\DesignDirectStochAlphaMed{3.2}
\newcommand\DesignDirectStochAlphaHigh{1.8}
\newcommand\DesignDirectStochAlphaLow{0.9}
\newcommand\DesignDirectStochZpeakMed{1.9}
\newcommand\DesignDirectStochZpeakHigh{1.6}
\newcommand\DesignDirectStochZpeakLow{1.1}

\graphicspath{{./fig/}}

\begin{document}

\title{Shouts and Murmurs: Combining Individual Gravitational-Wave Sources with the Stochastic Background to Measure the History of Binary Black Hole Mergers}
\correspondingauthor{Tom Callister}
\email{tcallister@flatironinstitute.org}

\author{Tom Callister}
\CCA

\author{Maya Fishbach}
\affiliation{Department of Astronomy and Astrophysics, University of Chicago, Chicago, IL 60637, USA}

\author{Daniel E. Holz}
\affiliation{Enrico Fermi Institute, Department of Physics, Department of Astronomy and Astrophysics, and Kavli Institute for Cosmological Physics, University of Chicago, Chicago, IL 60637, USA}

\author{Will M. Farr}
\CCA
\STBR

\date{\today}

\begin{abstract}
One of the goals of gravitational-wave astronomy is to quantify the evolution of the compact binary merger rate with redshift.
The redshift distribution of black hole mergers would offer considerable information about their evolutionary history, including their progenitor formation rate, the dependence of black hole formation on stellar metallicity, and the time delay distribution between formation and merger.
Efforts to measure the binary redshift distribution are currently limited, however, by the detection range of existing instruments, which can individually resolve compact binary merger events only out to $z\lesssim1$.
We present a novel strategy with which to measure the redshift distribution of binary black hole mergers well beyond the detection range of current instruments.
By synthesizing direct detections of individually resolved mergers with \textit{indirect} searches for the stochastic gravitational-wave background due to unresolved distant sources, we can glean information about the peak redshift, $z_p$, at which the binary black hole merger rate attains its maximum, even when this redshift is beyond the detection horizon.
Using data from Advanced LIGO and Virgo's first and second observing runs, we employ this strategy to place joint constraints on $z_p$ and the slope $\alpha$ with which the binary merger rate increases at low redshifts, ruling out merger rates that grow faster than $\alpha \gtrsim 7$ and peak beyond $z_p \gtrsim 1.5$.
Looking ahead, we project that approximately one year of observation with design-sensitivity Advanced LIGO will further break remaining degeneracies, enabling a direct measurement of the peak redshift of the binary black hole merger history.
\vspace{1cm}
\end{abstract}

\section{Introduction}

The Advanced LIGO~\citep{LIGO} and Advanced Virgo~\citep{Virgo} gravitational-wave experiments are rapidly transitioning between low- and high-statistics regimes.
With the LIGO-Virgo detections of eleven compact binary mergers during the past O1 and O2 observing runs~\citep{GWTC1} and tens more anticipated in the present O3 run~\citep{ObsProspects}, we can now begin to understand the ensemble properties of compact binaries, including the distributions of their component masses and spins~ \citep{Talbot2017,Farr2017,Fishbach2017,Talbot2018,Wysocki2019,Fishbach2019,O1O2populations}.
Beyond the distributions of these intrinsic binary parameters, we might also seek to understand the \textit{redshift distribution} of binary black hole mergers -- how the merger rate evolves as we look back to earlier times in the Universe's history.
If measured, the redshift distribution of compact binary mergers would offer substantial insight into the birth and evolution of compact binaries, encoding such properties as the time delay distribution between black hole formation and merger~\citep[see e.g.][]{2020arXiv200101025A}, the dependence of black hole production on stellar metallicity~\citep[see e.g.][]{2016Natur.534..512B}, and perhaps even the relative contributions from competing binary formation channels; e.g. field binaries, hierarchical triples, dynamical capture, or primordial black holes~\citep{DominikIII,Mandic2016,Mandel2018,Rodriguez2018}.

Study of the binary black hole redshift distribution, however, is made difficult by the limited range of existing gravitational-wave detectors.
Figure \ref{fig:rateModel}, for example, shows a typical model for the source-frame rate of binary black hole mergers as a function of redshift.
To obtain this figure, we assume progenitor formation following the star formation rate of~\cite{Madau2014} weighted by the fraction of stellar formation occurring at metallicities $Z\leq 0.3 Z_\odot$~\citep{Langer2006}.
We further adopt a $p(t_d) \propto t_d^{-1}$ probability distribution for the time delay $t_d$ between binary formation and merger, with $50\,\mathrm{Myr}\leq t_d \leq 13.5\,\mathrm{Gyr}$.
Within this simple model, the binary black hole merger rate peaks at $z\sim2$, while more sophisticated models generally predict merger rates peaking between redshifts $z\sim2$ to $4$, depending on the specific formation channel presumed~\citep{Dominik2013,Mapelli2017,Rodriguez2018,Baibhav2019,Santoliquido2020}.

In contrast, design-sensitivity Advanced LIGO is expected to successfully detect optimally-oriented $30+30\,M_\odot$ binary black holes only out to redshifts $z\lesssim1.2$~\citep{2017arXiv170908079C,ObsProspects}.
Current efforts to study the redshift distribution of compact binary mergers therefore attempt only to measure the leading-order, low-redshift evolution of the binary merger rate \citep{Fishbach2018,O1O2populations}; observation of the peak and subsequent turnover of the black hole redshift distribution is a challenge left to future third-generation detectors~\citep{Vitale2018}.

In this paper we demonstrate that present-day gravitational-wave observatories \textit{can} provide meaningful measurements of the high-redshift evolution of the compact binary merger rate.
We achieve these measurements by synthesizing the direct detections of compact binaries in the local Universe with an additional piece of information: the astrophysical stochastic gravitational-wave background~\citep{Romano2017,Christensen2019}.
Arising from the superposition of all distant individually-undetectable compact binaries, the stochastic gravitational-wave background manifests as excess correlated noise shared amongst a network of gravitational-wave detectors~\citep{Allen1999}.
The strength of the present-day gravitational-wave background is determined by the cumulative merger history of binary black holes, integrated across all redshifts~\citep{Phinney2001}.
The observation of (or even upper limits on) the gravitational-wave background can therefore be leveraged to place powerful constraints on the redshift distribution of binary mergers, complementary to those constraints gleaned from the direct detection of binaries in the local Universe (Sec.~\ref{sec:stoch-constraints}).

We apply our approach to existing data, finding that the synthesis of binary black hole detections \citep{GWTC1} and gravitational-wave background constraints~\citep{GW150914-implications,O1-Directional,O1-Isotropic,GW170817-implications,O2-Directional,O2-Isotropic} from Advanced LIGO and Advanced Virgo's first two observing runs already yields non-trivial constraints on the peak of the binary black hole redshift distribution (Sec.~\ref{sec:O1-O2}).
With additional data gathered from future observing runs, our method may enable a \textit{measurement} of this peak redshift within the next five years~(Sec.~\ref{sec:design}).

\section{High Redshift Constraints from the Gravitational-Wave Background}
\label{sec:stoch-constraints}

\begin{figure}
\includegraphics[width=0.47\textwidth]{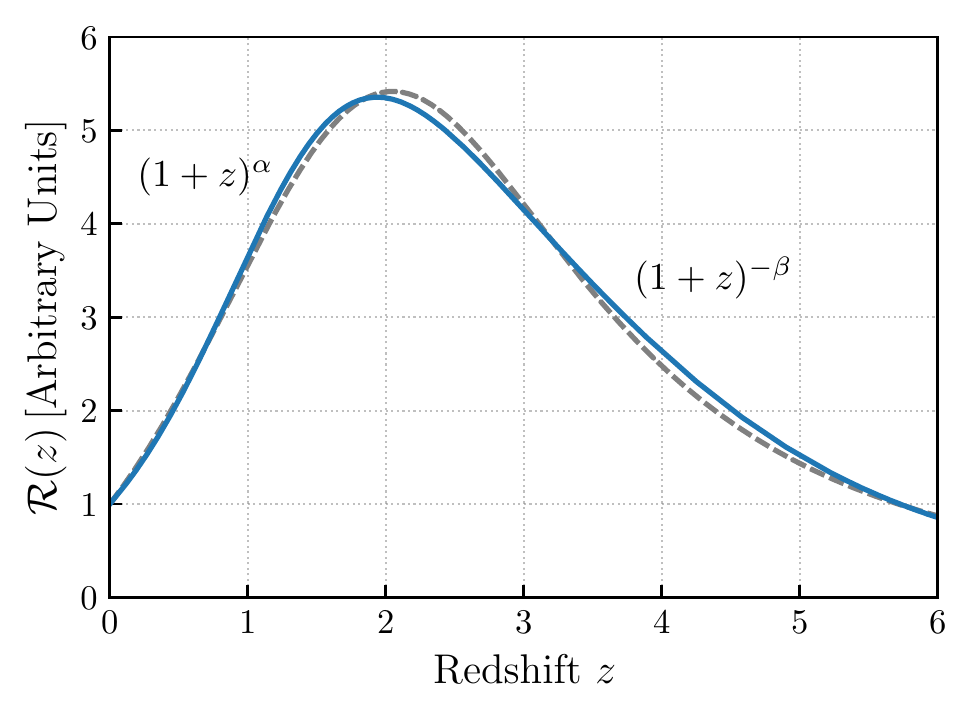}
\caption{Example prediction of the source-frame rate density of binary black hole mergers (solid blue), assuming progenitor formation that follows the rate of stellar formation at metallicities $Z\leq 0.3\,Z_\odot$~\citep{Langer2006,Madau2014}, and time delays $t_d$ between binary formation and merger distributed as $p(t_d)\propto t_d^{-1}$, with $0.05\,\mathrm{Gyr}\leq13.5\,\mathrm{Gyr}$.
In this work, we will adopt a phenomenological model for the binary black hole merger rate [Eq.~\eqref{eq:rate-density}] that allows for the same qualitative behavior as the prediction shown here, rising as $\mathcal{R}(z)\propto (1+z)^\alpha$ at $z\lesssim z_p$ and falling as $\mathcal{R}(z)\propto (1+z)^{-\beta}$ at redshifts $z\gtrsim z_p$.
The specific prediction plotted here, for instance, is well-fit by Eq.~\eqref{eq:rate-density} using $\alpha = 1.9$, $\beta = 3.4$, and $z_p = 2.4$ (the dashed grey curve).
}
\label{fig:rateModel}
\end{figure}

In their O1 and O2 observing runs, Advanced LIGO and Virgo confidently detected ten binary black hole mergers, the most distant of which (GW170729) may have occurred at $z\approx0.5$~\citep{GWTC1,Chatziioannou2019}.
Together, these ten events have recently allowed for the first exploration of the binary black hole merger rate's evolution with redshift.
Adopting a model
	\begin{equation}
	\mathcal{R}(z) = \mathcal{R}_0 \left(1+z\right)^\alpha,
	\end{equation}
for the source-frame merger rate per comoving volume \citep{Fishbach2018}, \cite{O1O2populations} find $\alpha = 6.5^{+9.1}_{-9.3}$ at 90\% credibility.
Thus, in the local universe, the binary black hole merger rate (probably) increases with redshift.

If the binary black holes observed with LIGO and Virgo are born from stellar progenitors, then the black hole merger rate cannot continue to increase out to arbitrarily high redshifts.
Instead, it must reach a maximum at some peak redshift, $z_p$, and then decay to zero as star formation ceases in the very early Universe.
Generically, we can describe this complete merger history with a phenomenological model of the form~\citep{Madau2014,Madau2017}:
	\begin{equation}
	\label{eq:rate-density}
	\mathcal{R}(z) = \mathcal{C}(\alpha,\beta,z_p) \frac{\mathcal{R}_0 \,\left(1+z\right)^\alpha}{1+\left(\frac{1+z}{1+z_p}\right)^{\alpha+\beta}}\,,
	\end{equation}
allowing a source-frame merger rate that initially evolves as $\mathcal{R}(z) \propto (1+z)^\alpha$, reaches a maximum near $z_p$, and subsequently falls as $\mathcal{R}(z)\propto (1+z)^{-\beta}$.
The example binary black hole merger rate shown in Fig.~\ref{fig:rateModel}, for example, is well-fit by this phenomenological model using $\alpha = 1.9$, $\beta = 3.4$, and $z_p = 2.4$, shown via a dashed grey curve.
The normalization constant $\mathcal{C}(\alpha,\beta,z_p) = 1+\left(1+z_p\right)^{-\alpha-\beta}$ ensures that $\mathcal{R}(0) = \mathcal{R}_0$.

At present the direct detection of binary black holes with Advanced LIGO and Virgo can offer no meaningful constraints on $z_p$ or $\beta$.
In O2, the range within which Advanced LIGO could detect a typical $30+30\,M_\odot$ binary black hole \citep[averaging over sky location and binary orientation; see][]{2017arXiv170908079C} was $z\lesssim0.5$; in the future O5 observing run this range may be pushed to $z\lesssim1.2$~\citep{ObsProspects}.
Meanwhile, if the black hole merger rate roughly follows the star formation rate, it should peak at $z_p\gtrsim2$, well beyond our ability to probe with direct detections.

We have another piece of information at our disposal, however.
Although individually undetectable, the superposition of \textit{all} distant binary black holes gives rise to a stochastic gravitational-wave background, detectable in the form of excess cross-power between widely-separated detectors~\citep{Romano2017,Christensen2019}.
The stochastic gravitational-wave background is conventionally described by a dimensionless energy-density spectrum~\citep{Allen1999}
	\begin{equation}
	\Omega(f) = \frac{1}{\rho_c} \frac{d\rho_\textsc{gw}}{d\ln f},
	\end{equation}
where $\frac{d\rho_\textsc{gw}}{d\ln f}$ is the present-day energy density in gravitational-waves per logarithmic frequency interval and $\rho_c = \frac{3 H_0^2 c^2}{8\pi G}$ is the Universe's critical energy density.
Here, $c$ is the speed of light, $G$ is Newton's constant, and $H_0$ is Hubble's constant; we adopt $H_0 = 70\,\mathrm{km}\,\mathrm{s}^{-1}\,\mathrm{Mpc}^{-1}$.

The energy density arising from the population of binary black hole mergers is given by~\citep{Phinney2001}:
	\begin{equation}
	\label{eq:Omg-zIntegral}
	\Omega(f) = \frac{f}{\rho_c} \int_0^{z_\mathrm{max}} dz \frac{\mathcal{R}(z) \left\langle\frac{dE_s}{df_s}|_{f(1+z)}\right\rangle}{(1+z) H(z)}.
	\end{equation}
Here, $\langle dE_s/df_s \rangle$ is the source-frame energy spectrum radiated by a single binary~\citep{Ajith2008}, averaged over the binary black hole population.
If the intrinsic parameters of individual binary black holes (e.g. their masses and spins) are denoted by $\phi$ and have distribution $p(\phi)$, then
	\begin{equation}
	\label{eq:average-energy}
	\left\langle \frac{dE_s}{df_s} \right\rangle = \int d\phi\,p(\phi) \frac{dE_s}{df_s}(\phi).
	\end{equation}
Note that in Eq.~\eqref{eq:Omg-zIntegral} we evaluate $\langle dE_s/df_s \rangle$ at the source-frame frequency $f(1+z)$.
Meanwhile, $H(z) = H_0 \sqrt{\Omega_\textsc{m}(1+z)^3 + \Omega_\Lambda}$ is the Hubble parameter at redshift $z$ (neglecting radiation density).
We take the energy densities of matter and dark energy to be $\Omega_\textsc{m} = 0.3$ and $\Omega_\Lambda = 0.7$, respectively.
Finally, the integral in Eq.~\eqref{eq:Omg-zIntegral} is taken up to a cutoff redshift $z_\mathrm{max}$; we fix $z_\mathrm{max} = 10$, beyond which we expect virtually no star formation and hence no black hole mergers (assuming stellar progenitors).
Alternatively, allowing $z_\mathrm{max}$ itself to vary as another free parameter may help to provide constraints on binary black holes of non-stellar origin, like the mergers of primordial black holes~\citep{Mandic2016,2016arXiv161008725W,2017PhRvL.119v1104K}.

The energy density, $\Omega(f)$, measured by stochastic searches is, in essence, a weighted integral over the binary black hole merger history $\mathcal{R}(z)$, sensitive to the \textit{total} number of past  mergers.
Thus, if the local rate $\mathcal{R}_0$ is independently fixed by direct detections, then knowledge of $\Omega(f)$ provides strong bounds on the possible values of $\alpha$, $\beta$, and $z_p$.
This is true even given a \textit{non-detection} of the gravitational-wave background.
To illustrate this, we can consider how the signal-to-noise ratio (S/N) of the gravitational-wave background varies with $\alpha$ and $z_p$.

Given a model $\Omega_M(f)$ for the true energy-density spectrum, the signal-to-noise ratio of the gravitational-wave background is~\citep{Allen1999}
	\begin{equation}
	\label{eq:stoch-snr}
	\mathrm{S/N} = \frac{\bigl(\hat C | \gamma\,\Omega_M\bigr)}{\sqrt{\left(\gamma\,\Omega_M | \gamma\,\Omega_M\right)}}.
	\end{equation}
Here,
	\begin{equation}
	\hat C(f) = \frac{1}{T} \frac{20\pi^2}{3 H_0^2} f^3 \tilde s_1(f) \tilde s^*_2(f)
	\end{equation}
is the cross-correlation statistic between the strains $\tilde s_1(f)$ and $\tilde s_2(f)$ measured by two gravitational-wave detectors~\citep{Romano2017,Callister2017}, and we have defined an inner product
	\begin{equation}
	\label{eq:stoch-innerProduct}
	\left(A|B\right) = 2 T \left(\frac{3H_0^2}{10\pi^2}\right)^2 \int_0^\infty df \frac{\tilde A(f) \tilde B^*(f)}{f^6 P_1(f) P_2(f)},
	\end{equation}
where $P_i(f)$ is the one-sided noise power spectral density of detector $i$ and $T$ is the total observation time.
In the presence of a gravitational-wave background, the expectation value of $\hat C(f)$ is
	\begin{equation}
	\label{eq:C-expectation}
	\langle \hat C(f) \rangle = \gamma(f) \Omega(f)
	\end{equation}
and its variance is $\langle \hat C(f) \hat C(f')\rangle = \delta(f-f')\sigma^2(f)$, with
	\begin{equation}
	\label{eq:C-std}
	\sigma^2(f) = \frac{1}{T} \left(\frac{10\pi^2}{3 H_0^2}\right)^2 f^6 P_1(f) P_2(f).
	\end{equation}
The factor $\gamma(f)$ in Eqs.~\eqref{eq:stoch-snr} and \eqref{eq:C-expectation}, known as the overlap reduction function, quantifies the geometrical sensitivity of a given detector pair to an isotropic gravitational-wave background~\citep{Christensen1992,Flanagan1993}.
The optimal S/N occurs when we choose a model $\Omega_M(f) = \Omega(f)$ matching the true energy density; the expected S/N in this case is~\citep{Allen1999}
	\begin{equation}
	\langle \mathrm{S/N}\rangle_\mathrm{opt}  = \sqrt{\left(\gamma\,\Omega | \gamma\,\Omega\right)}.
	\end{equation}

\begin{figure}
\includegraphics[width=0.47\textwidth]{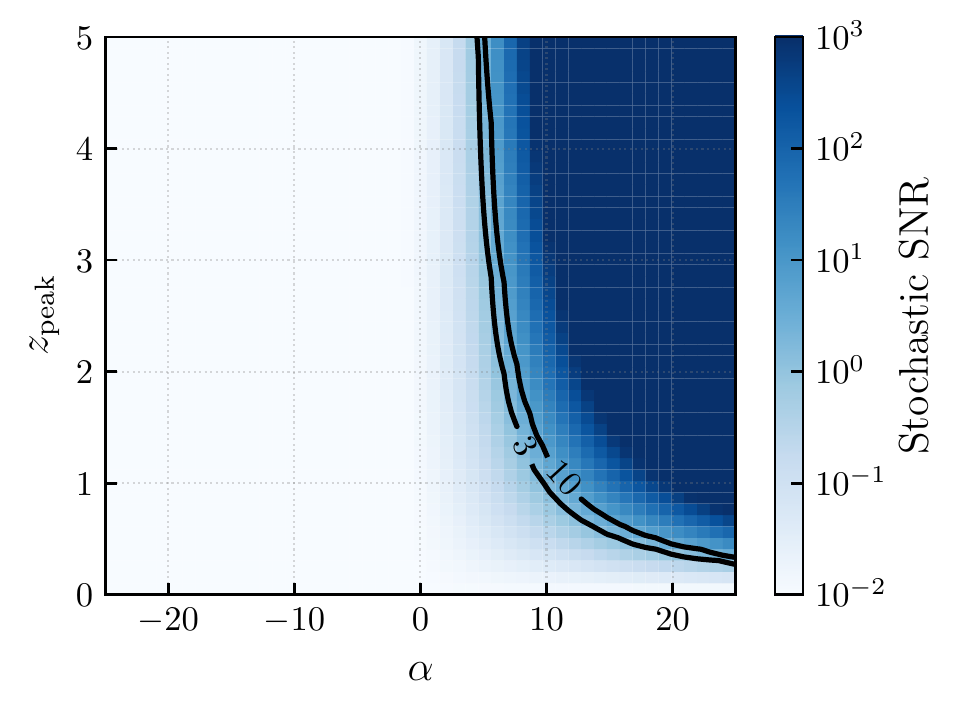}
\caption{Optimal signal-to-noise ratio with which the binary black hole stochastic background should be visible in Advanced LIGO's O1 and O2 observing runs, as a function of the leading slope, $\alpha$, and peak redshift, $z_p$, of the merger rate $\mathcal{R}(z)$; see Eq.~\eqref{eq:rate-density}.
For purposes of illustration, we have fixed $\mathcal{R}_0 = 30\,\mathrm{Gpc}^{-3}\,\mathrm{yr}^{-1}$ and $\beta = 3$, and assumed equal mass binaries with chirp masses $\mathcal{M}_c = 30\,M_\odot$.
The two black curves trace contours of constant signal-to-noise ratios, at $\langle\mathrm{S/N}\rangle_\mathrm{opt} = 3$ and $10$.
Given our choices of $\mathcal{R}_0$, $\beta$, and binary mass distribution, the non-detection of a stochastic gravitational-wave background in O1 and O2~\citep{O1-Isotropic,O2-Isotropic,Renzini2019} excludes values of $\alpha$ and $z_p$ at which $\mathrm{SNR}_\mathrm{opt} \gtrsim 3$, ruling out a large fraction of the $\alpha$--$z_p$ parameter space.
}
\label{fig:stochasticSNRs}
\end{figure}

In Fig.~\ref{fig:stochasticSNRs} we plot the optimal S/N with which the gravitational-wave background \textit{would} have appeared in O1 and O2 as a function of possible values for $\alpha$ and $z_p$.
In this example we fix $\mathcal{R}_0 = 30\,\mathrm{Gpc}^{-3}\,\mathrm{yr}^{-1}$ and $\beta = 3$, and assume a population of equal mass binaries with chirp mass $\mathcal{M}_c = 30\,M_\odot$.
If $\alpha\lesssim5$, virtually no stochastic signal is expected, consistent with the non-detection of the gravitational-wave background in O1~\citep{O1-Isotropic} and O2~\citep{O2-Isotropic,Renzini2019}.
However, the expected S/N rises sharply towards the upper-right corner of Fig.~\ref{fig:stochasticSNRs}.
In particular, if $\alpha\gtrsim 5$ and $z_p\gtrsim1$, we should have seen an extraordinarily loud stochastic gravitational-wave signal.
The fact that no such background was detected means that we can already reject this portion of parameter space, ruling out binary black hole backgrounds rising faster than $\alpha\sim5$ and peaking beyond $z_p\sim1$.
We note, though, that these exact limits depend strongly on the assumed local merger rate $\mathcal{R}_0$ and black hole mass distribution (and to a lesser extent on $\beta$), and so the results in Fig.~\ref{fig:stochasticSNRs} should be taken as an example only.
In Sect.~\ref{sec:O1-O2} below, we will instead seek to \textit{simultaneously} measure these different properties, leveraging both the observational limits on the stochastic gravitational-wave background and the current catalog of direct binary black hole detections.

So far, our argument has implicitly assumed that the distribution of binary black hole parameters is independent of redshift, such that the average energy radiated by a given binary [Eq.~\eqref{eq:average-energy}] does not vary with $z$.
This is not necessarily the case.
It is possible, for instance, that black holes born at high redshifts are preferentially more massive, due to the increased stellar masses predicted to occur at low metallicities~\citep{Belczynski2010,Spera2015,GW150914-astro}, although more recent work suggests that the mass distribution of \textit{merging} binaries may be approximately constant~\citep{Mapelli2019}.
By neglecting the possibility increased masses at higher redshifts, the constraints we obtain on $\alpha$ and $z_p$ are \textit{conservative}.
Given a fixed observational limit on $\Omega(f)$, any presumed increase in the average radiated energy $\langle dE_s/df_s \rangle$ must be balanced be a decrease in the merger rate $\mathcal{R}(z)$ at high redshifts, yielding stricter limits than those shown in Fig.~\ref{fig:stochasticSNRs}.
Nevertheless, one could incorporate effects like metallicity-dependent masses in this analysis by amending Eq.~\eqref{eq:average-energy} to additionally include integration over distributions of formation redshifts or progenitor metallicities~\citep{GW150914-implications,O1-Isotropic}.

\section{Peak Redshift Constraints from O1 and O2}
\label{sec:O1-O2}

The best constraints on $\mathcal{R}(z)$ will come from neither the direct detection of binary black holes nor the gravitational-wave background searches considered separately, but instead from a joint analysis that self-consistently synthesizes both sources of information.
In this paper we perform the first such joint analysis, synthesizing stochastic data and direct black hole observations to hierarchically measure the redshift distribution of binary black hole mergers.
We take as inputs the integrated cross-correlation spectrum $\hat C(f)$ measured between the LIGO Hanford and Livingston detectors~\citep{O1-Isotropic,O2-Isotropic} during O1 and O2, as well as parameter estimation results for each of the ten binary black hole mergers comprising the LIGO and Virgo GWTC-1 catalog~\citep{GWTC1}.

In order to robustly constrain $\mathcal{R}(z)$, it will also be important to simultaneously fit for the \textit{mass} distribution of binary black holes.
In Fig.~\ref{fig:stochasticSNRs} above, for instance, the exact exclusion region depends on our presumed black hole mass distribution: heavier or lighter black holes would increase or decrease the expected energy density $\Omega(f)$, leading us to draw different conclusions about $\mathcal{R}(z)$ in the case of a stochastic non-detection.
Strong degeneracies also exist between the inferred mass and redshift distributions of directly-detected black hole mergers~\citep{Fishbach2018}.
A dearth of detections at large redshifts, for example,  simply implies a low merger rate for high-mass binaries, since low-mass binaries go undetected at large distances.
This can be explained either by a low overall rate at high redshifts, or by a mass distribution that prefers low-mass binaries.

Consider a population of binary black hole mergers, with a local merger rate per unit comoving volume $\mathcal{R}_0$ and whose mass and redshift distributions are characterized by parameters $\Lambda$.
The likelihood of obtaining data $\{d_i\}_{i=1}^{N_\mathrm{obs}}$ from $N_\mathrm{obs}$ direct detections, as well as a stochastic cross-correlation spectrum $\hat C(f)$, is
	\begin{equation}
	p\big(\hat C,\{d_i\} | \Lambda,\mathcal{R}_0\big) =  p_\textsc{bbh}(\{d_i\} | \Lambda,\mathcal{R}_0) p_\mathrm{stoch}(\hat C | \Lambda,\mathcal{R}_0),
	\end{equation}
which has been factored into a direct-detection and a stochastic term.

The likelihood, $p_\textsc{bbh}(\{d_i\} | \Lambda,\mathcal{R}_0)$, of our direct binary black hole detections is given by~\citep{Loredo2004,Taylor2018,Mandel2019}
	\begin{equation}
	\label{eq:initial-cbc-likelihood}
	\begin{aligned}
	&p_\textsc{bbh}(\{d_i\}|\Lambda,\mathcal{R}_0) \\
	&\hspace{1.5cm}\propto\big[N(\Lambda,\mathcal{R}_0)\, \xi(\Lambda)\big]^{N_\mathrm{obs}} e^{-N(\Lambda,\mathcal{R}_0)\xi(\Lambda)} \\
	&\hspace{3.25cm}\times \prod_{i=1}^{N_\mathrm{obs}} \frac{\int p(d_i | \phi) p(\phi |\Lambda) d\phi}{\xi(\Lambda)}.
	\end{aligned}
	\end{equation}
Here, $p(d_i | \phi )$ is the likelihood for event $i$ given its component masses $m_1$ and $m_2$ and redshift $z$, together abbreviated as $\phi = \{m_1,m_2,z\}$.
Meanwhile, $p(\phi | \Lambda)$ is the ensemble distribution of these source parameters.
The quantity $N(\Lambda,\mathcal{R}_0)$ is the total number of binary black hole mergers (both observed and unobserved) expected to occur during our observation time; see Eq.~\eqref{eq:N-to-R} below.
Observational selection effects are captured by the factor $\xi(\Lambda)$, the fraction of all binary black holes that we expect to successfully detect.
If $P_\mathrm{det}(\phi)$ is the probability of successfully detecting an event with parameters $\phi$, then
	\begin{equation}
	\xi(\Lambda) = \int P_\mathrm{det}(\phi) p(\phi|\Lambda) d\phi.
	\end{equation}
In our analysis we precompute $P_\mathrm{det}(\phi)$ over a grid of masses and redshifts, using the semi-analytic prescription of \cite{Finn1993}, and requiring detections to have a matched filter signal-to-noise ratio of $\rho>8$ in a single detector.

In practice, we do not have direct access to the likelihoods, $p(d_i|\phi)$, needed to compute Eq.~\eqref{eq:initial-cbc-likelihood}.
Instead, we have discrete samples $\{\phi_{i}\}$ drawn from each event's \textit{posterior} distribution $p(\phi | d_i)$, obtained via parameter estimation with Monte Carlo integration or nested sampling~\citep{Veitch2015}.
Parameter estimation itself is performed while assuming some default prior, $p_\mathrm{pe}(\phi)$, that is generally \textit{not}\/ equal to the population prior $p(\phi|\Lambda)$ appearing in Eq.~\eqref{eq:initial-cbc-likelihood}.
To evaluate Eq.~\eqref{eq:initial-cbc-likelihood}, we must therefore replace the integral with an average over discrete samples, weighting each sample with $p_\mathrm{pe}^{-1}(\phi)$ to undo the influence of the prior used in parameter estimation:
	\begin{equation}
	\label{eq:hierarchical-likelihood-discrete}
	\begin{aligned}
	&p_\textsc{bbh}(\{d_i\}|\Lambda,\mathcal{R}_0) \\
	&\hspace{1.5cm}\propto \big[N(\Lambda,\mathcal{R}_0)\, \xi(\Lambda)\big]^{N_\mathrm{obs}} e^{-N(\Lambda,\mathcal{R}_0)\,\xi(\Lambda)} \\
		&\hspace{2.5cm}\times
		\prod_{i=1}^{N_\mathrm{obs}} \frac{1}{\xi(\Lambda)} \left\langle
		\frac{p(\phi_{i} | \Lambda)}{p_\mathrm{pe}(\phi_{i})}
		\right\rangle_\mathrm{samples}.
	\end{aligned}
	\end{equation}

The stochastic cross-correlation spectrum $\hat C(f)$, meanwhile, is generally obtained through the weighted combination of a large number of measurements performed over short $\mathcal{O}(100\,\mathrm{s})$ time segments~\citep{Allen1999,Romano2017}, and so the likelihood $p_\mathrm{stoch}(\hat C |\Lambda,\mathcal{R}_0)$ is well-approximated as a Gaussian~\citep{Mandic2012,Callister2017}:
	\begin{equation}
	\label{eq:stoch-likelihood}
	\begin{aligned}
	&p_\mathrm{stoch}(\hat C |\Lambda,\mathcal{R}_0) \\& \propto
	\exp\left[{- \frac{1}{2} \left(\hat C - \gamma\, \Omega_M(\Lambda,\mathcal{R}_0) | \hat C - \gamma\,\Omega_M(\Lambda,\mathcal{R}_0)\right)}\right],
	\end{aligned}
	\end{equation}
where $\Omega_M(\Lambda,\mathcal{R}_0;f)$ is our model energy-density spectrum and we have used the inner product defined in Eq.~\eqref{eq:stoch-innerProduct}.

We model the intrinsic redshift distribution of binary black hole mergers as
	\begin{equation}
	\label{eq:redshift-probability}
	p(z|\alpha,\beta,z_p) \propto \frac{1}{1+z} \mathcal{R}(\alpha,\beta,z_p;z) \frac{dV_c}{dz},
	\end{equation}
where $\mathcal{R}(\alpha,\beta,z_p;z)$ is given in Eq.~\eqref{eq:rate-density} and $\frac{dV_c}{dz}$ is the comoving volume per unit redshift;
note that Eq.~\eqref{eq:redshift-probability}, once normalized, is independent of the local merger rate $\mathcal{R}_0$.
The leading factor of $(1+z)^{-1}$ transforms between source-frame and detector-frame times.
Correspondingly, the total number $N$ of mergers expected to occur during our observation time $T$ is
	\begin{equation}
	\label{eq:N-to-R}
	N(\alpha,\beta,z_p,\mathcal{R}_0) = T \int_0^{z_\mathrm{max}}dz \,\frac{1}{1+z} \mathcal{R}(\alpha,\beta,z_p,\mathcal{R}_0;z) \frac{dV_c}{dz}.
	\end{equation}

Following \citet{Fishbach2018} and~\citet{O1O2populations}, we model the ensemble distribution of primary black hole masses as a power law
	\begin{equation}
	\label{eq:p-m1}
	p(m_1 | \kappa, M_\mathrm{min}, M_\mathrm{max}) \propto
	\begin{cases}
	m_1^{-\kappa} & (M_\mathrm{min}\leq m_1 \leq M_\mathrm{max}) \\
	0 & (\mathrm{else})
	\end{cases}
	\end{equation}
and assume a flat distribution
	\begin{equation}
	\label{eq:p-m2}
	p(m_2 | m_1, M_\mathrm{min}) =
	\begin{cases}
	\frac{1}{m_1 - M_\mathrm{min}} & (M_\mathrm{min}\leq m_2\leq m_1) \\
	0 & (\mathrm{else})
	\end{cases}
	\end{equation}
of secondary masses.

\begin{table}
\centering
\setlength{\tabcolsep}{6pt}
\renewcommand{\arraystretch}{0.9}
\caption{
Priors placed on the hyperparameters describing the binary black hole mass and redshift distributions; see Eqs.~\eqref{eq:redshift-probability}--\eqref{eq:p-m2}.
\vspace{-0.2cm}
}
\begin{tabular}{c | c | c c}
\hline
\hline
Parameter & Prior & Minimum & Maximum \\ [0.3ex]
\hline\hline
$\alpha$ & Uniform & -25 & 25 \\
$\beta$ & Uniform & 0 & 10 \\
$z_p$ & Uniform & 0 & 4 \\
$\mathcal{R}_0$ & Log-uniform & $10^{-1}$ & $10^{3}$ \\
$\kappa$ & Uniform & -4 & 12 \\
$ M_\mathrm{max}/M_\odot$ & Uniform & 30 & 100 \\[0.3ex]
\hline
\hline
\end{tabular}
\label{tab:priors}
\end{table}

\begin{figure*}
\centering
\includegraphics[width=0.45\textwidth]{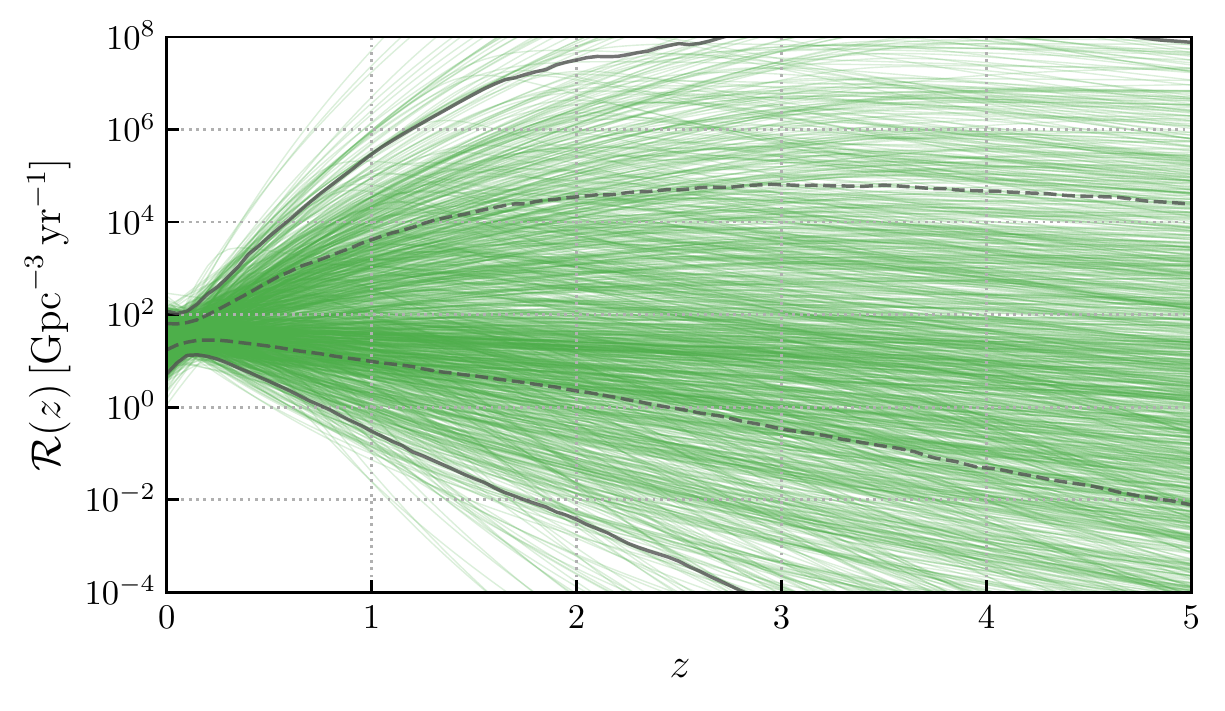} \hspace{0.5cm}
\includegraphics[width=0.45\textwidth]{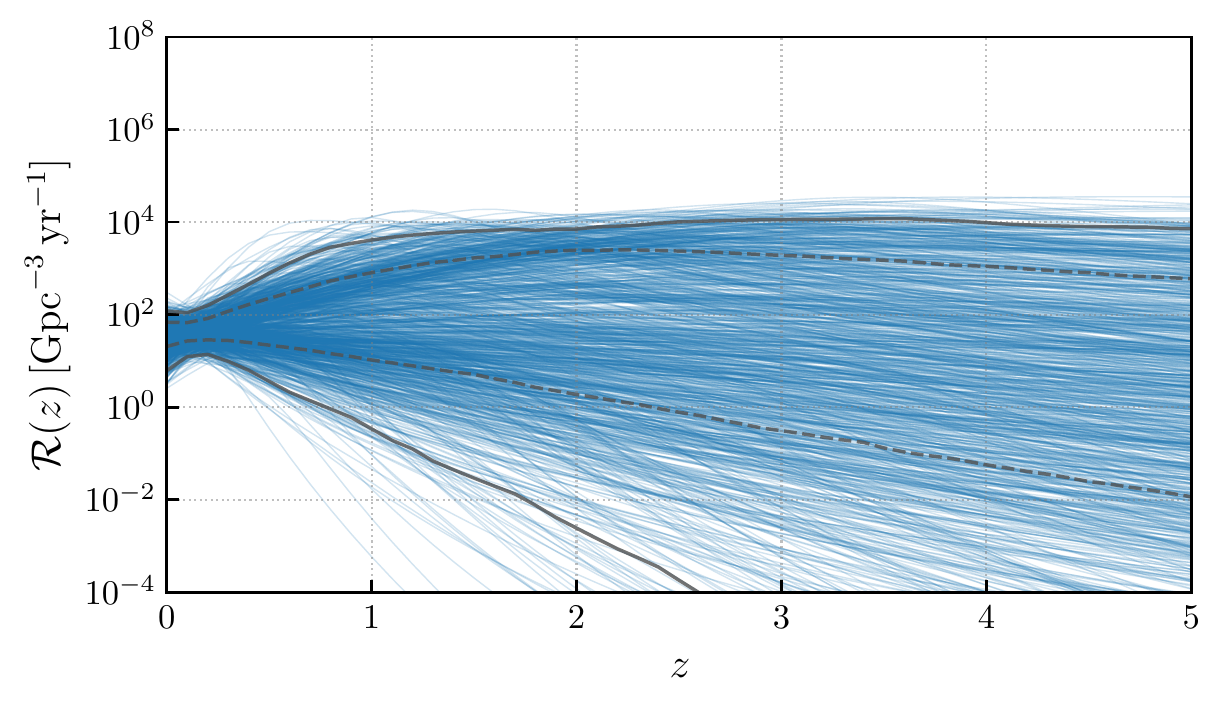}
\caption{
Posterior on the rate density $\mathcal{R}(z)$ of binary black hole mergers as a function of redshift, given the ten binary black holes comprising GWTC-1 (\textit{left}), and the joint analysis of these ten detections with O1 and O2 searches for the stochastic gravitational-wave background (\textit{right}).
The rate density is parameterized as in Eq.~\eqref{eq:rate-density}, and the dashed and solid grey curves show the central 68\% and 95\% credible bounds on $\mathcal{R}(z)$ at each redshift.
The direct GWTC-1 detections alone yield a measurement of the local merger rate and marginally constraint the slope $\alpha$ with which the rate evolves at low redshift (see also Fig.~\ref{fig:corner-CBC}), but give no constraints on the high-redshift behavior of $\mathcal{R}(z)$.
The non-detection of a stochastic gravitational-wave background in Advanced LIGO's O1 and O2 observing runs, meanwhile, imposes an upper limit on the net merger rate across all redshifts.
The joint analysis of direct detections and stochastic data can therefore exclude rate densities rising above $\mathcal{R}\gtrsim10^4\,\mathrm{Gpc}^{-3}\,\mathrm{yr}^{-1}$, placing joint constraints on $\alpha$ and the peak redshift $z_p$ at which $\mathcal{R}(z)$ reaches its maximum (see Fig.~\ref{fig:corner-CBCstoch}).
}
\label{fig:merger-rates}
\end{figure*}

\begin{figure*}
\centering
\includegraphics[width=0.7\textwidth]{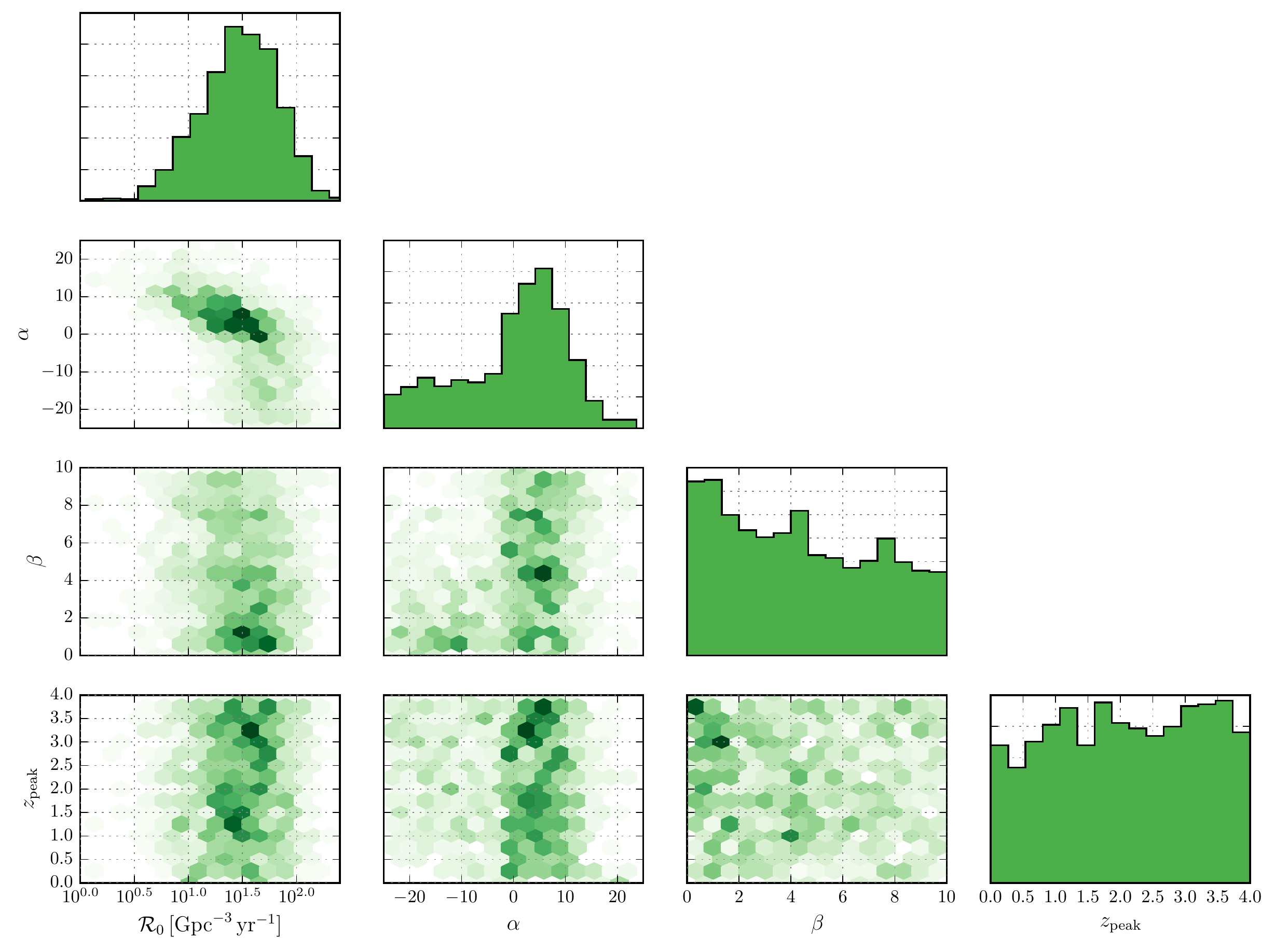}
\caption{
Posterior distribution on the local density $\mathcal{R}_0$, leading slope $\alpha$, trailing slope $\beta$, and peak redshift $z_p$ of the binary black hole merger rate $\mathcal{R}(z)$ [Eq.~\eqref{eq:rate-density}], given the ten binary black hole mergers comprising GWTC-1.
We have marginalized over the parameters $\kappa$ and $M_\mathrm{max}$ governing the black hole mass distribution [Eq.~\eqref{eq:p-m1}].
The GWTC-1 detections yield marginal constraints on $\alpha$, but offer no information about $z_p$ or $\beta$.
This posterior is used to construct the $\mathcal{R}(z)$ samples on the left side of Fig.~\ref{fig:merger-rates}.
Full parameter estimation results, including bounds on $\kappa$ and $M_\mathrm{max}$, are given in Table~\ref{tab:injected}.
}
\label{fig:corner-CBC}
\vspace{\floatsep}
\includegraphics[width=0.7\textwidth]{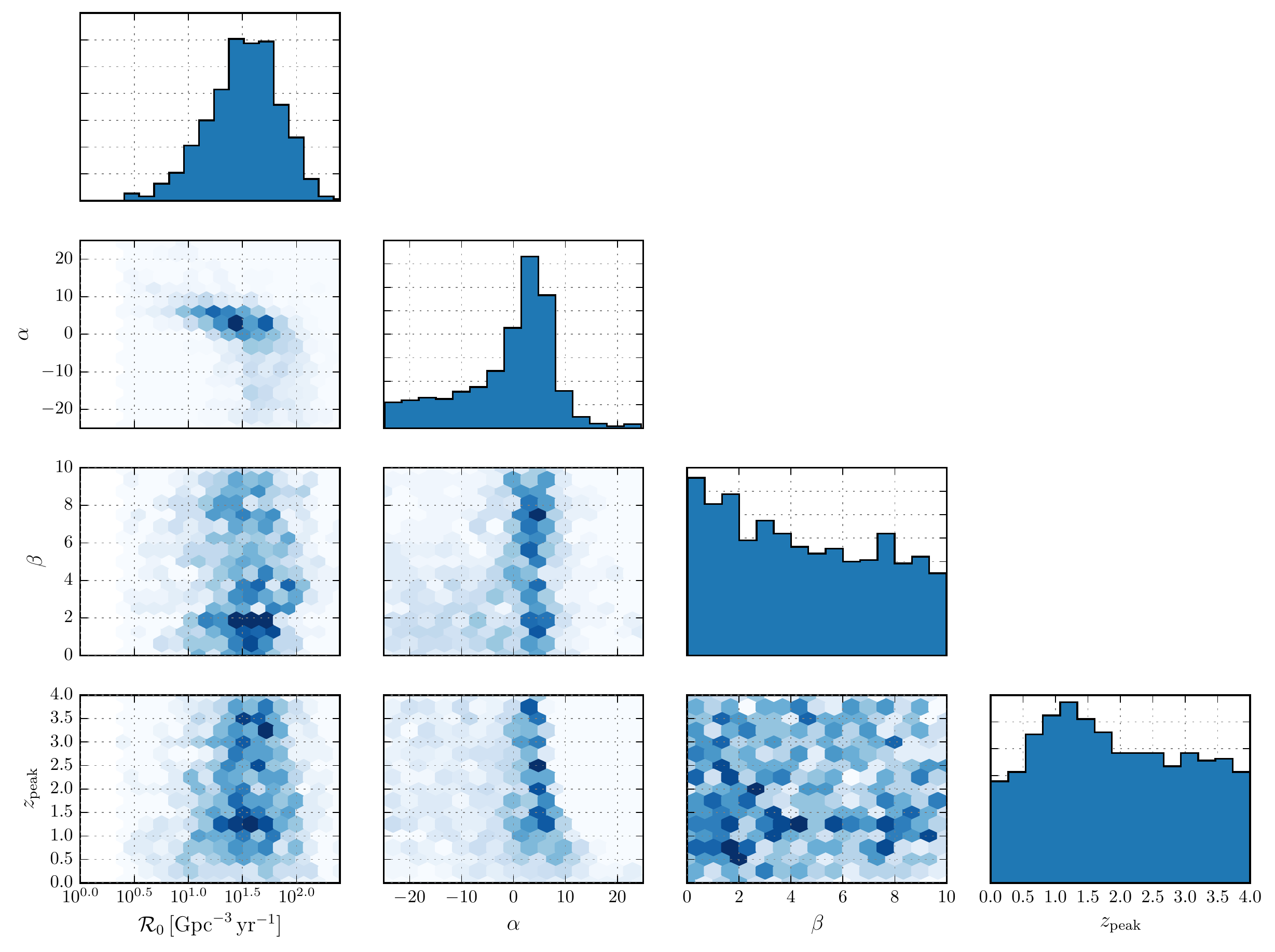}
\caption{
As in Fig.~\ref{fig:corner-CBC}, but incorporating a joint analysis using the GWTC-1 binary black holes as well as Advanced LIGO limits on the stochastic gravitational-wave background from O1 and O2.
Although the inclusion of stochastic measurements does not affect the marginalized one-dimensional posteriors, the non-detection of a gravitational-wave background by Advanced LIGO imposes a joint constraint on $\alpha$ and $z_p$, as seen in the lower-left subplot, ruling out rate densities that evolve faster than $\alpha\gtrsim7$ and reach maxima at redshifts beyond $z_p\gtrsim1$.
Draws from this posterior are used to generate the rate evolution constraints on the right side of Fig.~\ref{fig:merger-rates}.
}
\label{fig:corner-CBCstoch}
\end{figure*}

In our analysis we fix $M_\mathrm{min} = 5\,M_\odot$, while hierarchically inferring the parameters $\{\alpha,\beta,z_p,\mathcal{R}_0,\kappa,M_\mathrm{max}\}$ of the binary black hole redshift and mass distributions.
We adopt the priors listed in Table~\ref{tab:priors}, and perform inference using \texttt{emcee}~\citep{ForemanMackey2013}.
For every iteration of our sampler, we evaluate the direct-detection likelihood in Eq.~\eqref{eq:hierarchical-likelihood-discrete}, using Eq.~\eqref{eq:N-to-R} to convert the proposed event rate density $\mathcal{R}_0$ to a total number of mergers $N$.
We then compute a model stochastic energy-density spectrum, integrating over the proposed mass and redshift distributions [in Eqs.~\eqref{eq:average-energy} and Eq.~\eqref{eq:Omg-zIntegral}, respectively] of the binary black hole population, thereby evaluating the stochastic contribution [Eq.~\eqref{eq:stoch-likelihood}] to the overall likelihood.

Figure~\ref{fig:merger-rates} shows our resulting posterior on the rate evolution of binary black hole mergers, using the direct GWTC-1 detections alone (left) and combining direct detections with existing stochastic search results (right).
Each trace in these figures represents a draw from our $\{\alpha,\beta,z_p,\mathcal{R}_0\}$ posterior. The left panel of Fig.~\ref{fig:merger-rates} is directly comparable to Fig.~6 of \cite{O1O2populations}.
Figures~\ref{fig:corner-CBC} and \ref{fig:corner-CBCstoch} show the corresponding posteriors on these parameters, marginalized over $\kappa$ and $M_\mathrm{max}$.
Full parameter estimation results are listed in Table~\ref{tab:injected}.

Direct detections alone allow a measurement of the local merger rate to $\mathcal{R}_0 = \DirectRateMed^{+\DirectRateHigh}_{-\DirectRateLow}\,\mathrm{Gpc}^{-3}\,\mathrm{yr}^{-1}$ at 95\% credibility (the most precise measurement actually occurs at the ``waist'' seen at $z\sim0.1$). This is consistent with the results of~\cite{O1O2populations}.
Direct observations also allow us to roughly constrain $\alpha$, with a moderate preference for $\alpha\sim 5$ shown in Fig.~\ref{fig:corner-CBC}.
Significant uncertainties remain, however.
At 95\% credibility, we find $\alpha = \DirectAlphaMed^{+\DirectAlphaHigh}_{-\DirectAlphaLow}$, and, since the $\alpha$ posterior extends all the way to our lower prior bound, we can only robustly constrain $\alpha\leq\DirectAlphaUpperLim$.
Direct detections offer no information about $\beta$ or $z_p$.
Correspondingly, in Fig.~\ref{fig:merger-rates} we have virtually no constraints on the merger rate beyond $z\sim1$.
At $z=1.5$, for example, the local merger rate could plausibly lie anywhere between $10^{-4}$--$10^8\,\mathrm{Gpc}^{-3}\,\mathrm{yr}^{-1}$, a range spanning twelve orders of magnitude.

\begin{figure}
\centering
\includegraphics[width=0.45\textwidth]{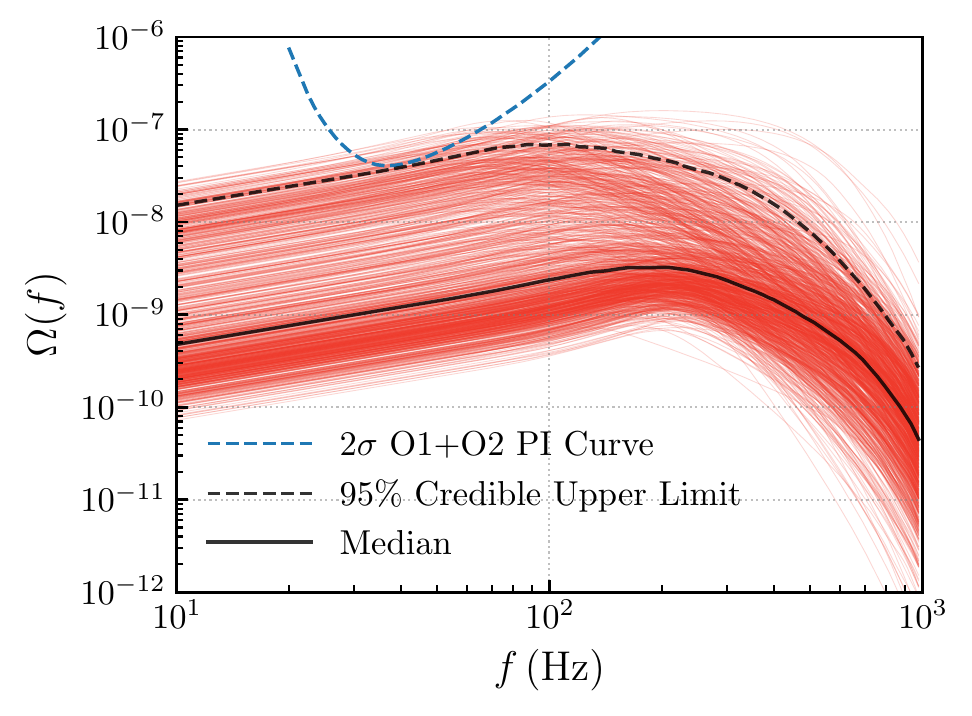}
\caption{
Posterior on the energy-density spectrum $\Omega(f)$ [see Eq.\eqref{eq:Omg-zIntegral}] of the binary black hole stochastic background, given the stochastic upper limits and direct binary black hole detections made by Advanced LIGO and Virgo during O1 and O2.
Each red trace corresponds to a posterior sample drawn from Fig.~\ref{fig:corner-CBCstoch}; the range of predictions shown here therefore incorporates our uncertainty in the mass and redshift distributions of binary black holes.
The solid and dashed black curves mark the median and $95\%$ credible upper limit on $\Omega(f)$, respectively.
For comparison, the dashed blue curve shows Advanced LIGO's $2\sigma$\,power-law integrated curve~\citep{Thrane2013} illustrating its sensitivity to the stochastic background following O2.
}
\label{fig:omega-gw}
\end{figure}

In contrast, the inclusion of O1 and O2 stochastic search data provides a hard upper bound on the high-redshift merger rate; our non-detection of the gravitational-wave background effectively excludes rate densities that rise above $\mathcal{R}(z)\gtrsim10^4\,\mathrm{Gpc}^{-3}\,\mathrm{yr}^{-1}$.
This additional constraint is reflected in Fig.~\ref{fig:corner-CBCstoch}.
While the inclusion of O1 and O2 stochastic data does not notably alter the one-dimensional marginal posteriors, it \textit{does} significantly alter our joint posterior on $\alpha$ and $z_p$.
As argued in Sec.~\ref{sec:stoch-constraints}, the non-detection of a stochastic gravitational-wave signal provides a joint constraint on these two parameters, rejecting a large portion of the $\alpha-z_p$ parameter space.
When this stochastic exclusion region is combined with the constraint on $\alpha$ from direct GWTC-1 detections, we can already see hints of a preferred contour in the $\alpha$--$z_p$ plane.

Although the primary goal of this analysis is to measure the evolution of the binary black hole merger rate, it additionally provides a self-consistent framework for predicting the energy density $\Omega(f)$ of the binary black hole background using both the known population properties of local binary black holes and upper limits from Advanced LIGO and Advanced Virgo stochastic searches~\citep{O1-Isotropic,O2-Isotropic}.
For every posterior sample in Fig.~\ref{fig:corner-CBCstoch} (including the mass parameters $M_\mathrm{max}$ and $\kappa$ not shown there) we compute the corresponding binary black hole energy density using Eq.~\eqref{eq:Omg-zIntegral}.
The result, shown in Fig.~\ref{fig:omega-gw}, is a prediction for the binary black hole stochastic background that is marginalized over our uncertainty in both the mass distribution and rate evolution of binary black holes, and subject to the measured upper limits from Advanced LIGO.

Within Fig.~\ref{fig:omega-gw}, the dashed black curve traces the 95\% credible upper limit on $\Omega(f)$ at each frequency.
For comparison, the dashed blue curve shows the $2\sigma$\,``power-law integrated (PI) curve''~\citep{Thrane2013} quantifying Advanced LIGO's integrated sensitivity to the gravitational-wave background following O1 and O2; energy-density spectra lying above this curve will generally be observed with $\mathrm{S/N}\geq 2$.
As expected, the 95\% credible limit on $\Omega(f)$ lies nearly tangent to the PI curve.
The solid black curve, meanwhile, marks the median predicted energy-density.
At 25\,Hz, this median prediction gives $\Omega(25\,\mathrm{Hz}) = 8.8\times10^{-10}$, comparable to the prediction made by \cite{O2-Isotropic}: $\Omega(25\,\mathrm{Hz}) = 5.3\times10^{-10}$.
The \textit{uncertainty} on our predicted energy-density spectrum, though, is considerably larger.
While the \cite{O2-Isotropic} model includes uncertainty on the local rate density $\mathcal{R}_0$ of binary black hole mergers, it makes stringent assumptions concerning the subsequent evolution of the merger rate with redshift, assumptions that carry considerable systematic uncertainty.
In contrast, Fig.~\ref{fig:omega-gw} includes marginalization over all possible redshift distributions, making this systematic uncertainty explicit.

\section{Advanced LIGO at Design Sensitivity}
\label{sec:design}

The continued synthesis of direct detections with stochastic search results will offer increasingly strong information regarding the leading slope, $\alpha$, and peak, $z_p$, of the binary black hole merger history.
Additional binary black holes detected in the local Universe will yield ever tighter posteriors on $\alpha$, while continued time integration by stochastic searches will reject a growing fraction of the joint $\alpha$--$z_p$ posterior space.
Eventually these two effects will meet, converging to produce a true measurement of both $\alpha$ and $z_p$.

To illustrate this, here we anticipate the results that will soon be possible with design-sensitivity Advanced LIGO.
We simulate a mock catalog of 500 binary black hole detections, drawn from a population whose mass distribution is characterized by $\kappa=1.2$, $M_\mathrm{max} =45\,M_\odot$, and $M_\mathrm{min}=5\,M_\odot$.
We assume a redshift distribution given by $\alpha=3$, $\beta=3$, $z_p=2$, and $\mathcal{R}_0=30\,\mathrm{Gpc}^{-3}\,\mathrm{yr}^{-1}$.
With this choice of local merger rate, we would expect to detect these 500 binary black holes after $T\sim1.2$ years of observation with design-sensitivity Advanced LIGO.

\begin{figure*}
\centering
\includegraphics[width=0.45\textwidth]{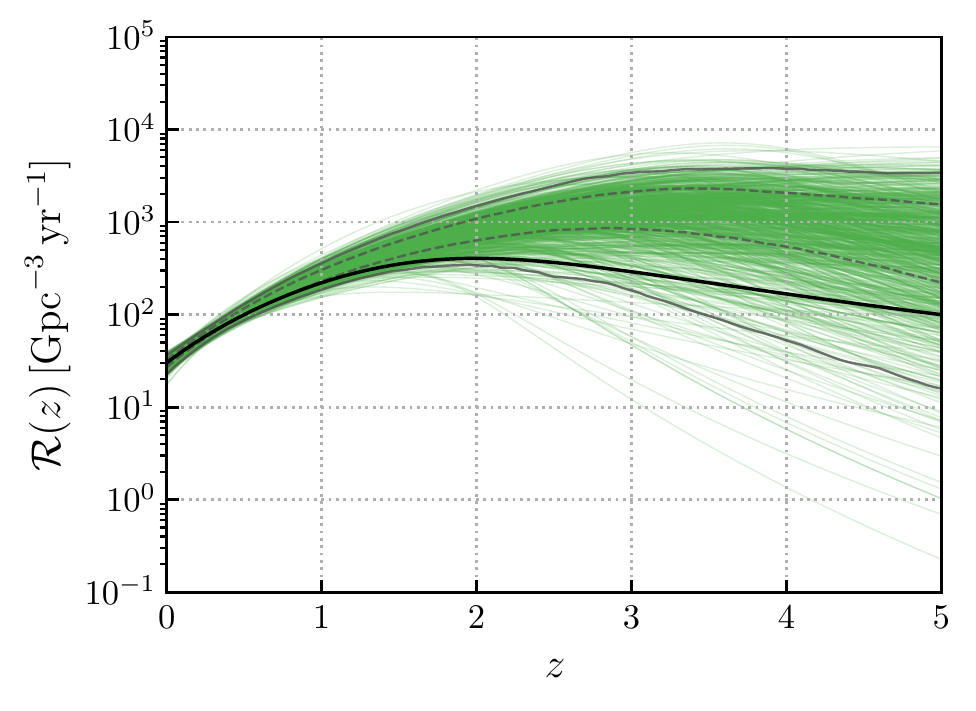} \hspace{0.5cm}
\includegraphics[width=0.45\textwidth]{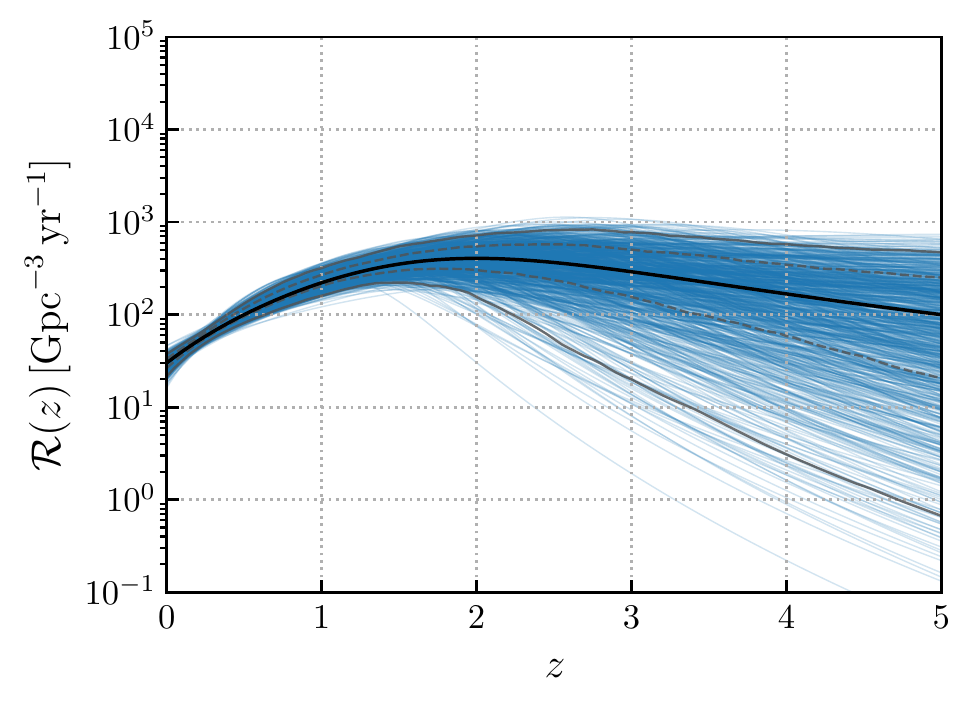}
\caption{
Expected posterior on the rate density $\mathcal{R}(z)$ of binary black hole mergers, given 1.2 years of observation with design-sensitivity Advanced LIGO.
We analyze a catalog of 500 mock detections as well as simulated measurements of the stochastic gravitational-wave background.
The left subplot (green) shows results obtained from mock detections alone, while the right subplot shows results given by the synthesis of mock detections with gravitational-wave background measurements.
In each case, the dashed and solid grey curves show our 68\% and 95\% credible symmetric bounds on the merger rate evolution, and the black trace shows the ``true'' injected merger rate.
Although the peak of this merger rate occurs at $z_p = 2$, well beyond Advanced LIGO's horizon, the joint analysis of direct detections with stochastic data allows us to reconstruct $\mathcal{R}(z)$, yielding the posteriors shown in Fig~\ref{fig:corner-design}.
}
\label{fig:merger-rates-design}
\end{figure*}
\begin{figure*}
\centering
\includegraphics[width=0.75\textwidth]{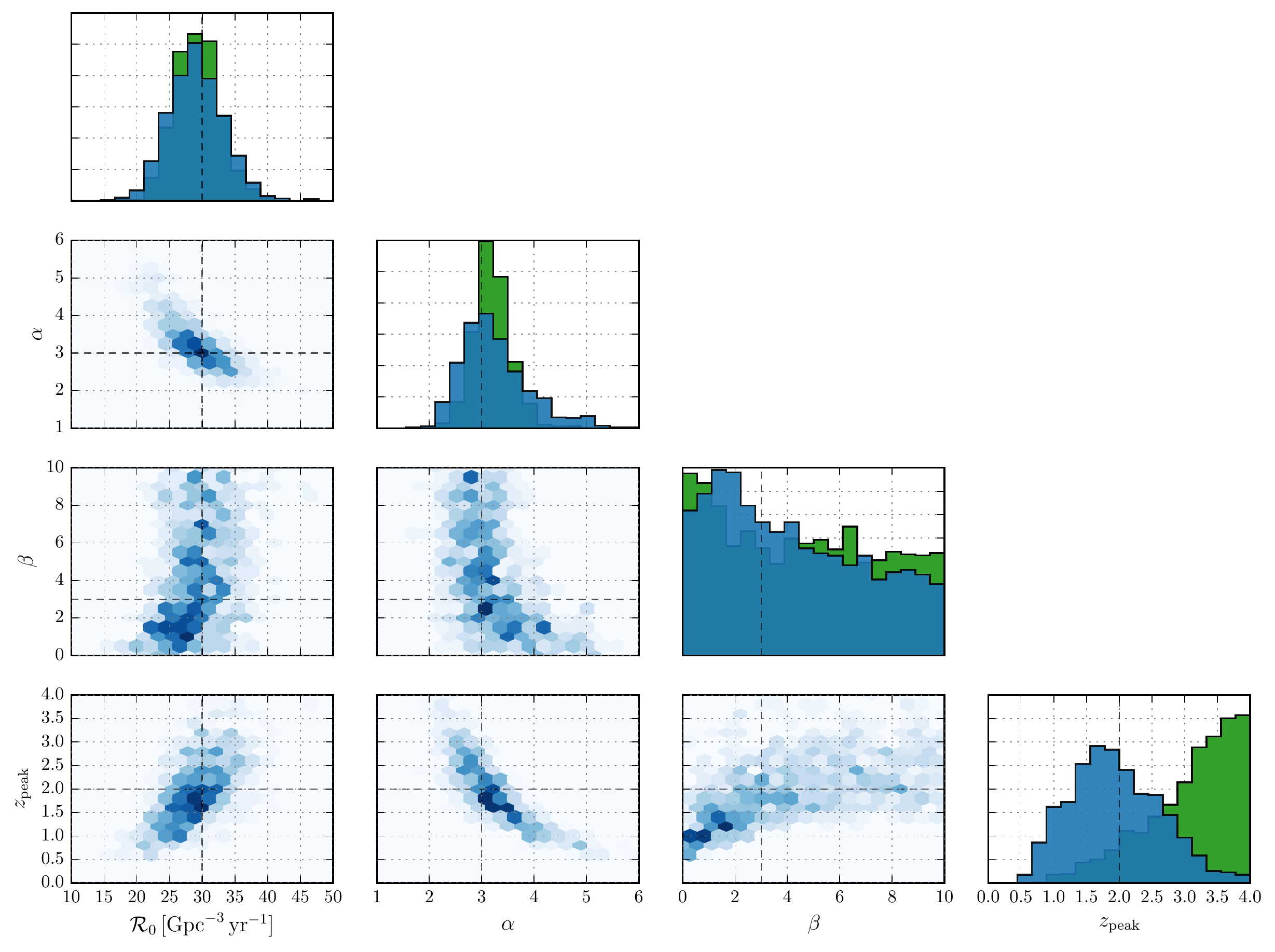}
\caption{
Expected posteriors on the local density $\mathcal{R}_0$, leading slope $\alpha$, trailing slope $\beta$, and peak redshift $z_p$ of the binary black hole merger rate after 1.2 years of Advanced LIGO observation at design sensitivity.
The green marginal distributions correspond to the left-hand side of Fig.~\ref{fig:merger-rates-design}, obtained using a mock catalog of direct BBH detections; blue distributions (both one- and two-dimensional) correspond to the right-hand side of Fig.~\ref{fig:merger-rates-design}, given by the synthesis of the BBH catalog with simulated stochastic measurements.
We have marginalized over the parameters $\kappa$ and $M_\mathrm{max}$ characterizing the black hole mass distribution.
The catalog of synthetic detections provides reasonable measurements of $\mathcal{R}_0$ and $\alpha$, but offers only a lower bound on $z_p$.
The addition of stochastic search results imposes an upper bound on $z_p$; taken together, we bound $z_p = \DesignDirectStochZpeakMed^{+\DesignDirectStochZpeakHigh}_{-\DesignDirectStochZpeakLow}$ at 95\% credibility.
}
\label{fig:corner-design}
\end{figure*}

We construct our mock catalogue following \cite{Fishbach2018}.
For each event, we draw an ``observed'' signal-to-noise ratio
	\begin{equation}
	\rho_\mathrm{obs} \sim \mathcal{N}(\rho,1)
	\end{equation}
from a Gaussian distribution about the event's true signal-to-noise ratio $\rho$, calculated in a detector with a noise power spectral density given by the Advanced LIGO ``design sensitivity" curve of~\cite{ObsProspects}.
We require our detected events to have $\rho_\mathrm{obs}>8$ in a single detector.
For each detected event, we draw an observed maximum-likelihood chirp mass
	\begin{equation}
	\log \mathcal{M}_{\mathrm{obs}} \sim \mathcal{N}\left(\log\mathcal{M} ,\sigma_\mathcal{M}\frac{8}{\rho_\mathrm{obs}}\right)
	\end{equation}
and symmetric mass ratio
	\begin{equation}
	\eta_\mathrm{obs} \sim \mathcal{N}\left(\eta ,\sigma_\eta \frac{8}{\rho_\mathrm{obs}}\right),
	\end{equation}
where $\mathcal{M}$ and $\eta$ are the event's true parameters and we adopt characteristic uncertainties $\sigma_\mathcal{M} = 0.08$ and $\sigma_\eta = 0.022$.
We then draw synthetic likelihood samples about $\log \mathcal{M}_{\mathrm{obs}}$ and $\eta_\mathrm{obs}$, with variances consistent with the above distributions. This prescription gives realistic uncertainties on the measured component masses and distances of BBH detections, matching the typical uncertainties reported in~\cite{Vitale2017}.

We encapsulate a binary's inclination angle and sky location in a single \cite{Finn1993} projection factor $\Theta$, which quantifies a signal's amplitude reduction due to suboptimal viewing angles and/or sky placement.
If $\rho_\mathrm{opt}$ is a binary's optimal signal-to-noise ratio (i.e. face-on and directly overhead), then $\Theta \rho_\mathrm{opt}$ is the event's actual signal-to-noise ratio.
For each mock event, we draw a maximum-likelihood projection factor from
	\begin{equation}
	\Theta_\mathrm{obs}\sim \mathcal{N}\left(\Theta,\sigma_\Theta \frac{8}{\rho_\mathrm{obs}}\right)
	\end{equation}
where $\sigma_\Theta = 0.15$, about which we draw likelihood samples $\{\Theta\}$.

Realistic redshift samples will be strongly correlated with an event's recovered S/N as well as its projection factor $\Theta$.
To capture these correlations, we first draw S/N ratio samples
	\begin{equation}
	\{\rho\} \sim \mathcal{N}\left(\rho_\mathrm{obs},1\right).
	\end{equation}
Then, noting that $\rho$ is inversely proportional to an event's luminosity distance $D_L$, we convert $\{\rho\}$ and $\{\Theta\}$ into luminosity distance samples via
	\begin{equation}
	\label{eq:dl}
	\frac{\{D_L\}}{1\,\mathrm{Gpc}} =  \rho_\mathrm{opt}(1\,\mathrm{Gpc})\frac{\{\Theta\}}{\{\rho\}},
	\end{equation}
where $\rho_\mathrm{opt}(1\,\mathrm{Gpc})$ is the binary's optimal signal-to-noise ratio at $1\,\mathrm{Gpc}$.

We additionally simulate cross-correlation measurements of the corresponding stochastic gravitational-wave background, assuming $T=1.2$ years of integration with Advanced LIGO's Hanford-Livingston baseline.
Our simulated cross-correlation spectra are drawn from
	\begin{equation}
	\hat C(f) \sim  \mathcal{N}\Bigl(\gamma(f)\Omega(f),\,\sigma(f)\Bigr),
	\end{equation}
where the gravitational-wave background's energy density $\Omega(f)$ is calculated using Eq.~\eqref{eq:Omg-zIntegral} and $\sigma(f)$ is given by Eq.~\eqref{eq:C-std}.
Given the binary black hole mass and redshift distributions assumed above and a 1.2 year integration time, the gravitational-wave background has amplitude $\Omega_0 = 2.2\times10^{-9}$ at $f=25\,\mathrm{Hz}$ and $\langle\mathrm{S/N}\rangle_\mathrm{opt}=4.2$.
In our particular noise realization, the binary black hole background is observed with $\mathrm{S/N} = 3.6$, representing a marginal detection.

Figure~\ref{fig:merger-rates-design} illustrates the posterior we obtain on $\mathcal{R}(z)$ using our simulated direct detections (left) and direct detections plus stochastic data (right).
Figure~\ref{fig:corner-design} shows the posterior on $\mathcal{R}_0$, $\alpha$, $\beta$, and $z_p$ for this latter case; as before, we have marginalized over the parameters governing the black hole mass distribution.
For reference, Fig.~\ref{fig:corner-design} also includes the one-dimensional marginalized posteriors obtained by direct detections alone (in green).
Full parameter estimation results for each case are given in Table~\ref{tab:injected}.

With 500 direct detections we can very precisely measure $\alpha=\DesignDirectAlphaMed^{+\DesignDirectAlphaHigh}_{-\DesignDirectAlphaLow}$ at 95\% credibility, yielding a tight fit to $\mathcal{R}(z)$ out to $z\sim1$.
By virtue of not directly observing a turnover of $\mathcal{R}(z)$, we can now place a lower limit $z_p\geq\DesignDirectZpeakLowerLim$.
Otherwise, we are again limited by Advanced LIGO's finite detection range.
The joint analysis of our direct detections and stochastic data, meanwhile, yields a qualitatively different picture.
Although the S/N of our simulated detection of the gravitational-wave background is somewhat marginal, it provides enough complementary information to rule out large $z_p$.
While the absolute merger rate remains uncertain at large redshifts, this future data would confidently measure $z_p = \DesignDirectStochZpeakMed^{+\DesignDirectStochZpeakHigh}_{-\DesignDirectStochZpeakLow}$.

\begin{table*}
\setlength{\tabcolsep}{8pt}
\renewcommand{\arraystretch}{1.1}
\caption{
95\% credible constraints on parameters governing the mass and redshift distribution of binary black hole mergers.
The second and third lines show true results given data from Advanced LIGO's O1 and O2 observing runs, using direct observations of binary mergers alone, as well as the synthesis of direct detections with constraints on the stochastic gravitational-wave background.
While both the ``Direct'' and ``Direct/Stochastic'' O1 \& O2 analyses give similar one-dimensional results, the inclusion of stochastic data excludes a non-trivial portion of the joint $\alpha-z_p$ space; see Fig.~\ref{fig:corner-CBCstoch}.
In neither case can we measure $\alpha$; instead we place an upper limit.
The fourth and fifth lines give parameter estimation results from our mock catalog corresponding to one year of Advanced LIGO observation at design sensitivity.
When analyzing mock direct detections alone, we can at best place a lower limit on the peak redshift $z_p$, while the inclusion of simulated stochastic data allows us to directly measure $z_p$.
None of the four cases give informative marginalized measurements of $\beta$, and so this parameter is excluded from the table.
\vspace{0.2cm}
}
\begin{tabular}{l | r r r r r}
\hline
\hline
Run & $M_\mathrm{max}\,[M_\odot]$ & $\kappa$ & $\alpha$ & $z_p$ & $\mathcal{R}_0\,[\mathrm{Gpc}^{-3}\,\mathrm{yr}^{-1}]$ \\
\hline
O1-O2: Direct
	& $\DirectMmaxMed^{+\DirectMmaxHigh}_{-\DirectMmaxLow}$
	& $\DirectGammaMed^{+\DirectGammaHigh}_{-\DirectGammaLow}$
	& $\leq\DirectAlphaUpperLim$
	& ---
	& $\DirectRateMed^{+\DirectRateHigh}_{-\DirectRateLow}$
	\\
O1-O2: Direct/Stochastic
	& $\DirectStochMmaxMed^{+\DirectStochMmaxHigh}_{-\DirectStochMmaxLow}$
	& $\DirectStochGammaMed^{+\DirectStochGammaHigh}_{-\DirectStochGammaLow}$
	& $\leq\DirectStochAlphaUpperLim$
	& ---
	& $\DirectStochRateMed^{+\DirectStochRateHigh}_{-\DirectStochRateLow}$
	\\
\hline
Design (Mock): Direct
	& $\DesignDirectMmaxMed^{+\DesignDirectMmaxHigh}_{-\DesignDirectMmaxLow}$
	& $\DesignDirectGammaMed^{+\DesignDirectGammaHigh}_{-\DesignDirectGammaLow}$
	& $\DesignDirectAlphaMed^{+\DesignDirectAlphaHigh}_{-\DesignDirectAlphaLow}$
	& $\geq \DesignDirectZpeakLowerLim$
	& $\DesignDirectRateMed^{+\DesignDirectRateHigh}_{-\DesignDirectRateLow}$
	\\
Design (Mock): Direct/Stochastic
	& $\DesignDirectStochMmaxMed^{+\DesignDirectStochMmaxHigh}_{-\DesignDirectStochMmaxLow}$
	& $\DesignDirectStochGammaMed^{+\DesignDirectStochGammaHigh}_{-\DesignDirectStochGammaLow}$
	& $\DesignDirectStochAlphaMed^{+\DesignDirectStochAlphaHigh}_{-\DesignDirectStochAlphaLow}$
	& $\DesignDirectStochZpeakMed^{+\DesignDirectStochZpeakHigh}_{-\DesignDirectStochZpeakLow}$
	& $\DesignDirectStochRateMed^{+\DesignDirectStochRateHigh}_{-\DesignDirectStochRateLow}$
	\\
\hline
\hline
\end{tabular}
\vspace{0.5cm}
\label{tab:injected}
\end{table*}

\section{Conclusions}

We present a powerful new constraint on the binary black hole redshift distribution, with implications for stellar evolution, black hole formation, and binary black hole formation and evolution.
By combining detections of compact binaries in the local Universe with measurements of (or upper limits on) the stochastic gravitational-wave background, we demonstrate that it is possible to explore the binary black hole redshift distribution at redshifts well beyond the present horizon of direct detections.
Using existing observations from the Advanced LIGO/Virgo O1 and O2 observing runs, we have obtained novel joint constraints on the low-redshift slope $\alpha$ and peak $z_p$ of the binary black hole merger rate [see Eq.~\eqref{eq:rate-density}].
In particular, we can reject merger rates that grow faster than $\alpha \gtrsim 7$ and peak beyond $z_p \gtrsim 1.5$.
These constraints will significantly improve with continued observation.
Given an approximately year-long observation period with design-sensitivity Advanced LIGO, we have demonstrated the possibility of \textit{directly measuring} $z_p$.

Although we have taken adopted a decidedly phenomenological model for the merger rate $\mathcal{R}(z)$ in this work, this is not the only possible approach.
If, for instance, one were willing to assume that binary black hole \textit{formation} is tied directly to the (potentially metallicity-dependent) star formation rate, as in Fig.~\ref{fig:rateModel}, one could instead seek to parametrize and measure the metallicity distribution of binary progenitors and the time delay distribution between binary formation and merger.

Looking ahead, future proposed ground-based gravitational-wave detectors like Cosmic Explorer and Voyager may be able to directly measure the rate of binary black hole mergers out to $z\gtrsim 10$~\citep{Vitale2018}.
However, even a more limited ability to explore the history of binary black hole mergers with present-day instruments will allow us to ask, sooner rather than later, questions of considerable astrophysical importance:
What are the progenitors of compact binary mergers, and when did they form?
What is the mean time delay between binary formation and merger?
How do black hole mergers across cosmic time connect to the evolution of stars and galaxies in the Universe?
The combination of individually resolved sources and the unresolved stochastic gravitational-wave background may soon provide answers.

\section*{Acknowledgements}

We would like to thank Nelson Christensen, Andrew Matas, and others within the LIGO Scientific Collaboration and Virgo Collaboration for helpful comments and conversation.
We additionally thank the anonymous referee, whose questions and feedback greatly improved the quality of this work.
TC and WMF thank the Simons Foundation for its generous support of the Flatiron Institute.
TC was also partially supported by the Josephine de Karman Fellowship Trust.
MF was supported by the NSF Graduate Research Fellowship Program under grant DGE-1746045. MF and DEH were supported by  NSF grant PHY-1708081.
They were also supported by the Kavli Institute for Cosmological Physics at the University of Chicago through an endowment from the Kavli Foundation.
DEH also gratefully acknowledges support from the Marion and Stuart Rice Award.
This research has made use of data, software and/or web tools obtained from the Gravitational Wave Open Science Center (https://www.gw-openscience.org), a service of LIGO Laboratory, the LIGO Scientific Collaboration and the Virgo Collaboration.
LIGO is funded by the U.S. National Science Foundation. Virgo is funded by the French Centre National de Recherche Scientifique (CNRS), the Italian Istituto Nazionale della Fisica Nucleare (INFN) and the Dutch Nikhef, with contributions by Polish and Hungarian institutes.


\bibliographystyle{aasjournal.bst}
\bibliography{Refs}

\end{document}